\documentclass[aps,showpacs,amsmath,amssymb,10pt, twocolumn]{revtex4}
\usepackage[latin9]{inputenc}
\usepackage{ifsym}
\usepackage{tipa}
\usepackage{amsmath}
\makeatletter

\newcommand{\lyxmathsym}[1]{\ifmmode\begingroup\def\b@ld{bold}
  \text{\ifx\math@version\b@ld\bfseries\fi#1}\endgroup\else#1\fi}

\@ifundefined{textcolor}{}
{%
 \definecolor{BLACK}{gray}{0}
 \definecolor{WHITE}{gray}{1}
 \definecolor{RED}{rgb}{1,0,0}
 \definecolor{GREEN}{rgb}{0,1,0}
 \definecolor{BLUE}{rgb}{0,0,1}
 \definecolor{CYAN}{cmyk}{1,0,0,0}
 \definecolor{MAGENTA}{cmyk}{0,1,0,0}
 \definecolor{YELLOW}{cmyk}{0,0,1,0}
}

\usepackage{latexsym}
\usepackage{bm}

\def\HollowBox #1#2{{\dimen0=#1 \advance\dimen0 by -#2
       \dimen1=#1 \advance\dimen1 by #2
        \vrule height #1 depth #2 width #2
        \vrule height 0pt depth #2 width #1
        \llap{\vrule height #1 depth -\dimen0 width \dimen1} 
       \hskip -#2
       \vrule height #1 depth #2 width #2}}

\makeatother

\begin{document}

\title{Bulk and Boundary Unitary Gravity in 3D: MMG$_2$}

\author{Bayram Tekin }

\email{btekin@metu.edu.tr}

\affiliation{Department of Physics,\\
 Middle East Technical University, 06800, Ankara, Turkey}

\date{\today}
\begin{abstract}
We construct a  massive spin-2 theory in 2+1 dimensions that is immune to the bulk-boundary unitarity conflict in anti-de Sitter space and hence amenable to holography. The theory is an extension of Topologically Massive Gravity, just like the recently found Minimal Massive Gravity (MMG), but it has two massive helicity modes instead of a single one. The theory admits all the solutions of TMG with a redefined topological parameter. We calculate the Shapiro time-delay and show that flat-space (local) causality is not violated. We show that  there is an interesting  relation between the theory we present here (which we call MMG$_2$), MMG and the earlier New Massive Gravity (NMG): Namely,  field equations of these theories are non-trivially related. We study the bulk excitations and boundary charges of the conformal field theory that could be dual to gravity. We also find the chiral gravity limit for which one of the massive modes becomes massless. The virtue of the model is that one does not have to go to the chiral limit to achieve unitarity in the bulk and on the boundary and the log-terms that appear in the chiral limit and cause instability do not exist in the generic theory.
\end{abstract}
\maketitle

\section{Introduction}
Gravity in $2+1$ dimensions has a counterintuitive richness: On the one side, naive counting from the metric leads to the conclusion that once gauge invariance (diffeomorphism invariance)  is taken into account, no local propagating  degrees of freedom (gravitons or gravity waves) exist, as in the case of Einstein's gravity. On the other hand, modifications of the theory, such as with higher powers of curvature introduce non-trivial local dynamics along with, usually, massive gravitons.  Therefore, while it is very hard to get non-linear, unitary, ghost-free massive gravity in 3+1 dimensions with 5 degrees of freedom, it is embarrassingly easy to get massive gravity with 2 degrees of freedom in one lower dimensions.  By now there there are several 3D models : Topologically Massive Gravity (TMG),\cite{DJT}, New Massive Gravity (NMG)\cite{NMG} and the recent Minimal Massive Gravity (MMG) \cite{Townsend1,Townsend2}. TMG is a parity-violating theory with a single spin-2 degree of freedom, NMG has a massive spin-2 excitation with both helicities, while MMG has a single massive parity-violating spin-2 excitation (same as TMG) but free of the bulk-boundary unitarity conflict in anti-de Sitter(AdS)spacetime that inflicts NMG and TMG. Ultimately, of course, research in 3D gravity aims at understanding or building  "quantum gravity"  in the physically relevant spacetime. For this purpose, obtaining a  unitary, nontrivial gravity theory that has a well-defined unitary conformal field theory on the boundary is an important step. For example, NMG which has 2 massive bulk excitations just like General Relativity in 4D (with a massless graviton) does not have a unitary conformal field theory (CFT) on the boundary. [Elaborate  extensions  of NMG could not resolve the issue \cite{Sinha,BINMG}.] Hence MMG stands alone as a curious non-trivial case of a 3D gravity which potentially has a viable boundary  CFT. Here we shall construct another theory that has this property and that propagates both helicities, albeit with different masses.

This work was inspired by \cite{Townsend1} and aimed  to build a 3D gravity with 2 massive helicity-2  modes (instead of the single one in MMG) that is free of the bulk-boundary unitarity conflict.The construction led to interesting connections between the existing massive gravity theories  and their chiral limits \cite{Strominger}. The field equations of MMG  read 
\begin{equation} 
G_{\mu\nu}+\Lambda_{0}\, g_{\mu\nu}+\frac{1}{\mu}C_{\mu\nu}+\frac{\gamma}{\mu^{2}}J_{\mu\nu}=0\,,\label{mmg_denk}
\end{equation}
where we have set the coefficent of the Einstein tensor to one. The Cotton and the $J$-tensors are  given in terms of the Schouten tensor, $S_{\mu\nu}=R_{\mu\nu}-\frac{1}{4}g_{\mu\nu}R$, as
\begin{equation}
C_{\mu\nu}=\eta_{\mu}\,^{\alpha\beta}\nabla_{\alpha}S_{\beta\nu},\hskip 0.5cm  J^{\mu\nu}\equiv \frac{1}{2}\eta^{\mu\rho\sigma}\eta^{\nu\alpha\beta}S_{\rho\alpha}S_{\sigma\beta}\, , 
\end{equation}
where $\eta^{\mu \nu\sigma}$ is the totally antisymmetric tensor.
As noted in \cite{Townsend1,Tekin,Alishahiha}, unitarity ranges of the bulk excitations with  $M_{g}^{2}=\mu^{2}\left ( 1+\frac{\gamma}{2\mu^{2}l^{2}}\right)^{2}+\frac{1}{l^{2}}$; $\Lambda = -1/\ell^2$ and the unitarity ranges of the boundary CFT's central charges 
\begin{equation}
c_{R/L}=\frac{3\text{\ensuremath{l}}}{2G_{3}}\left(1+\frac{\gamma}{2\mu^{2}\text{\ensuremath{l}}^{2}}\pm\frac{1}{\mu\text{\ensuremath{l}}}\right),
\end{equation}
are compatible. The $J$-tensor introduced in \cite{Townsend1}, while keeping TMG's bulk properties intact, makes the boundary theory unitary.  It also has the following non-zero covariant divergence 
\begin{equation} 
\nabla_{\mu}J^{\mu\nu}=\eta^{\nu \rho \sigma}S_{\sigma}{}^{\tau}C_{\rho\tau},
\label{div_J_tensor}
\end{equation}
which vanishes for the solutions of the theory.  The question we ask is the following: Is there another two-tensor that has a similar on-shell vanishing divergence in 3D that can be used to deform TMG to have two spin-2 degrees of freedom ?  Posed this way, the answer seems somewhat hard to get, but  the dimensions in the problem give us a hint. Next we construct this tensor.
\section{The new Tensor: $H_{\mu \nu}$}
Keeping in mind that we would like to deform TMG in such a way that we keep its healthier bulk properties intact, yet with a doubled number of excitations, we try the following tensor
\begin{equation}
H^{ \mu \nu }  \equiv \frac{1}{2}\eta^{\mu \alpha \beta }\nabla_\alpha C^\nu_\beta +  \frac{1}{2}\eta^{\nu \alpha \beta }\nabla_\alpha C^\mu_\beta 
\end{equation}
It is easy to show that the divergence of the $H$-tensor  is exactly minus that of the $J$-tensor 
\begin{equation}
\nabla_{\mu} H^{\mu \nu} = - \nabla_{\mu} J^{\mu \nu} = \eta^{\nu\rho\sigma}S_{\rho}{}^{\tau}C_{\sigma\tau}, \label{div}
\end{equation}
which is apriori quite unexpected since these two tensors are quite different. For example, their explicit forms read
\begin{eqnarray}
&&J_{\mu\nu}  =-S_{\mu}^{\rho}S_{\rho\nu}+ S S_{\mu\nu}+\frac{1}{2}g_{\mu\nu}\Big (S_{\rho\sigma}S^{\rho\sigma}-S^{2} \Big ), \\ \nonumber
&&H^{\mu \nu } =\Box S^{\mu \nu}- \nabla^\mu \nabla^\nu S + g^{\mu \nu}S_{\rho\sigma}S^{\rho\sigma} - 3 S^\mu_\alpha S^{\alpha \nu},  \label{H_tensor}
\end{eqnarray}
where $S \equiv g^{\mu \nu}S_{\mu \nu}$. Before we build our theory by deforming TMG with $H_{\mu \nu}$, let us look at some properties of this tensor: First of all it is traceless $g^{\mu \nu} H_{\mu \nu}=0$,  secondly, for all {\it solutions} of TMG, that is for $\gamma =0$ in (\ref{mmg_denk}), the $H$-tensor reduces to the Cotton tensor :
\begin{equation}
H_{\mu \nu} = -\mu C_{\mu \nu}.
\end{equation}
Thirdly, as a consequence of  (\ref{div}) one has the following Bianchi identity valid for all smooth metrics
\begin{equation}
\nabla_\mu \Big (  J^{\mu \nu} + H^{ \mu \nu} \Big ) =0,
\end{equation}
which then implies that the sum of the two tensors come from the variation of an action. Denoting 
the sum as $ K^{\mu \nu} \equiv   J^{\mu \nu} + H^{ \mu \nu} $, one can show that this action is nothing but the quadratic part of the NMG action \cite{Townsend1} given as
\begin{equation}
I  =  \int d^{3}x\,\sqrt{-g} \Big  ( R_{\mu\nu} R^{\mu \nu}  -\frac{3}{8} R^{2} \Big ), 
\end{equation}
which is a rather remarkable result. So clearly, MMG is built on "part" of the  NMG equations in such a way that on-shell Bianchi identity is satisfied, even though the field equations of MMG do not come from the variation of an action with the metric being the only field. In general one can deform TMG by adding to its field equation the two tensor $ a_1 J_{\mu \nu} + a_2 H_{\mu \nu}$; for $a_1 = a_2 $ one has the NMG deformed TMG which is Generalized Massive Gravity \cite{Sun}. Proper combination of $a_i$ also gives the theory studied in \cite{Setare}.

The above observation leads one to study the extension of TMG with the "sister" of the $J$-tensor, that is the $H$-tensor. Not to clutter the notation, let us not introduce any sign-adjusting parameters and study the following equations
\begin{equation} 
G_{\mu\nu}+\Lambda \, g_{\mu\nu}+\frac{1}{\mu}C_{\mu\nu} -\frac{1}{m^2} H_{\mu\nu}=0, \,\label{mmg2_denk}
\end{equation}
which has a unique maximally symmetric,  (A)dS vacuum  with $R = 6 \Lambda$ (just like TMG) since the $H$-tensor is traceless.  Observe that all the solutions of TMG also solve this theory with the slight modification that one should shift the topological mass as  
\begin{equation}
\mu \rightarrow  \frac{\mu}{2} \Big ( 1 \pm \sqrt{1 + \frac{4 \mu^2}{ m^2}} \Big ).
\end{equation}
Unlike the case of MMG, $\nabla_{\mu} H^{\mu \nu}$ does not vanish for all solutions but it does so for  a large class of solutions, including all the metrics that solve TMG \cite{sezgin, pope} such as algebraic Types $O,N,D$ and some Kundt solutions \cite{sezgin,emel} and many more:  For example, for all solutions of the form $H_{\mu \nu} =  f_1  S_{\mu \nu} + f_2 S_{\mu \rho} S^{\rho}_\nu$, where  $f_1$ and $f_2$ are scalars built on the curvature and they are not necessarily constant, but one has the condition that $H_{\mu  \nu}$ is traceless.  In general, among the solutions of   (\ref{mmg_denk}), one should take only the ones that satisfy $\eta^{\nu \rho \sigma}S^\tau_\sigma \square S_{\rho \tau}=0$, which restricts the solution space to TMG plus a large class of solutions including MMG's solutions when  $H_{\mu \nu} = \tilde{J}_{\mu \nu}$, where tilde refers to the traceless part.  As we are particularly interested in a holographically better behaved extension of TMG with two massive gravitons, (\ref{mmg2_denk}) is a good candidate  supplemented with the constraint  $\eta^{\nu \rho \sigma}S^\tau_\sigma \square S_{\rho \tau}=0$, which we now explore more.
\section{Graviton Spectrum}
Let us now find the particle spectrum about the (A)dS vacuum. For this purpose let us  rewrite the $H$-tensor as 
\begin{eqnarray}
H^{\mu \nu } &=&\Box {\mathcal{G}} ^{\mu \nu}+ (g^{\mu \nu} \Box -\nabla^\mu \nabla^\nu )S + g^{\mu \nu} {\mathcal{G}} _{\alpha \beta}^2 - 3 {\mathcal{G}} ^\mu_\alpha {\mathcal{G}} ^{\alpha \nu}  \nonumber \\
&&+ 2(S-\Lambda) {\mathcal{G}}  g^{\mu \nu} -6(S-\Lambda) {\mathcal{G}}^{\mu \nu},
\label{Hten}
\end{eqnarray}
with the cosmological Einstein as ${\mathcal{G}}_{\mu \nu} = G_{\mu \nu} + \Lambda  g_{\mu \nu}$ and ${\mathcal{G}} = -\frac{1}{2} ( R- 6 \Lambda)$. We have not yet used the field equation $R = 6 \Lambda$.
Linearization of  (\ref{Hten}) about the AdS background yields
\begin{equation}
(H_{\mu \nu} )_L = \bar\square {\mathcal{G}}^L_{\mu \nu}+(\bar g_{\mu \nu} \bar\square -\bar\nabla^\mu \bar \nabla^\nu )S_L  -2\Lambda \bar g_{\mu \nu} S_L - 3 \Lambda  {\mathcal{G}}^L_{\mu \nu}.
\end{equation} 
And since for all solutions, $S_L=0$, the linearized field equations become
\begin{equation}
\Big ( -\frac{1}{m^2}\ \delta^\beta_\mu \bar \square + \frac{1}{\mu} \eta_{\mu}\,^{\alpha \beta}\bar \nabla_\alpha +  (1 + \frac{3\Lambda}{m^2}) \delta^\beta_\mu \Big )  {\mathcal{G}}^L_{\beta \nu} =0.
\label{lineq}
\end{equation}
Since the linearized Einstein tensor is background diffeomorphism invariant, following \cite{Strominger}, we can choose the transverse traceless gauge which reduces the the linearized (cosmological) Einstein tensor to $ {\mathcal{G}}^L_{\mu \nu}  =  -\frac{1}{2} ( \bar \square - 2 \Lambda ) h_{\mu \nu} $  and the field equations to 
\begin{equation}
\Big ( \bar \square - 2 \Lambda \Big ) \Big ( -\frac{1}{m^2} \delta^\beta_\mu \bar \square + \frac{1}{\mu} \eta_{\mu}\,^{\alpha \beta}\bar \nabla_\alpha +  (1 + \frac{3\Lambda}{m^2})\delta^\beta_\mu \Big )  h_{\beta \nu} =0
\label{lin_h}
\end{equation}
Following \cite{Strominger,Sun} we can define four mutually commuting operators 
\begin{eqnarray}
&& ( \mathcal{D}^{L/R})_\mu\,^\nu = \delta_\mu^\nu \pm \ell \eta_{\mu}\,^{\alpha \nu} \bar\nabla_\alpha, \nonumber \\
&&( \mathcal{D}^{p_i})_\mu\,^\nu = \delta_\mu^\nu + \frac{1}{p_i} \eta_{\mu}\,^{\alpha \nu} \bar\nabla_\alpha, \,\,\, i =1,2,
\end{eqnarray}
which can be used to rewrite (\ref{lineq}) as  
\begin{equation}
\Big ( \mathcal{D}^L  \mathcal{D}^R   \mathcal{D}^{p_1}   \mathcal{D}^{p_2} h \Big )_{\mu \nu}=0.
\label{product}
\end{equation}
Form this construction, one can find that the $p$ parameters satisfy
\begin{equation}
p_1 + p_2 = -\frac{m^2}{\mu}, \hskip 1 cm  p_1  p_2 = -m^2
\label{pdenk1}
\end{equation}
with the solutions 
\begin{equation}
p_{1,2}= -\frac{ m^2}{ 2 \mu}  \pm \sqrt{ m^2 + \frac{m^4}{ 4\mu^2}} .
\label{pdenk2}
\end{equation}
We must now relate the $p$-parameters to the actual masses of the gravitons. For this purpose we can carry out two computations the first one is : $( \mathcal{D}^L  \mathcal{D}^R h )_{\mu\nu}  =0$, which says that the massless modes in AdS satisfy $ (\bar  \square  + \frac{2}{\ell^2} ) h_{\mu \nu} =0$, as one already knows from the linearized Einstein's theory.  The second one is 
\begin{equation}
( \mathcal{D}^{-p} \mathcal{D}^p h )_{\mu\nu} = -\frac{1}{p^2} ( \bar\square + \frac{3}{\ell^2} -p^2 )h_{\mu \nu} =0,
\end{equation}
which says that  two massive gravitons have the following masses
\begin{equation}
m_{i}^2 = p_i^2  -\frac{1}{\ell^2}.
\end{equation}
Asuming $m^2 \ge 0$, or  $ m^2 \le - 4 \mu^2$, Breiteinlohner-Freedman \cite{BF} (BF) bound in AdS, $ m_i^2  \ge  -\frac{1}{\ell^2}$, is satisfied, ensuring the non-tachyonic nature of these excitations.  Now our task is to compute the left and right central charges of this parity-violating theory. This can be done by following the  Brown-Hennaux procedure and boundary conditions \cite{BH} or a slightly modified version of  the conserved charge computation given in \cite{DT1,DT2} ( See also, \cite {Sun}.)
This somewhat lengthy computation yields
\begin{equation}
c_{R/L }= \frac{ 3 \ell}{2 G}  \Big ( 1 - \frac{1}{ 2 m^2 l^2 -1} \pm \frac{ 2 m^2 \ell^2}{ \mu \ell ( 2 m^2 \ell^2 -1)} \Big ),
\end{equation}
which reduce to those of TMG as $m^2 \rightarrow \infty$. Both charges are positive for 
\begin{equation}
\mu \ell  \ge  \frac{m^2 \ell^2}{ m^2 \ell^2 -1}.
\label{positivity}
\end{equation}
This is the unitarity condition on the boundary theory. When the bound is saturated, $c_L=0$  and the boundary theory is chiral analogous to the pure TMG case \cite{Strominger}.  At the chiral point, the right central charge reads 
\begin{equation}
c_R = \frac{ 6\ell}{G}  \frac{( m^2 l^2 -1 )}{2 m^2 \ell^2 -1},
\end{equation}
hence  $m^2 \ell^2 -1 \ge 0$ for $c_R \ge 0$.
At the chiral point, $p_1 \ell =1$ and $p_2 = -m^2 \ell$  and so the mases of the bulk excitations become 
\begin{equation}
m_1  = 0,  \hskip 1 cm  m_2^2 = m^4 \ell^2 -\frac{1}{\ell^2},
\end{equation}
Again, following the conserved charge construction given in {\cite{DT1,DT2} and {\cite{Tekin} as adjusted to the case at hand, one can compute the energy (with mass parameter $M$) and the angular momentum (with rotation parameter $a$)  of the BTZ black hole or any spacetime that asymptotes to such a solution as
\begin{equation}
E = \frac{1}{G} \Big ((1 - \frac{1}{m^2 \ell^2}) M  - \frac{a }{ \mu \ell^2} \Big), \hskip 0.3 cm 
J= \frac{1}{G} \Big ((1 - \frac{1}{m^2 \ell^2}) a  - \frac{M}{ \mu} \Big).
\end{equation}
Observe that at the chiral point, both $E$ and $J$ vanish, as is the case in TMG \cite{Olmez}. 
\section{Solutions of the Linearized Equations}
At a generic point, since the four operators commute in (\ref{product}), the most general solution can be written as 
\begin{equation}
h_{\mu \nu} = h_{\mu \nu}^L + h_{\mu \nu}^R + h_{\mu \nu}^{m_1} + h_{\mu \nu}^{m_2} ,
\end{equation}
each part satisfying the corresponding linear equation as 
\begin{equation}
{(\mathcal{D} h )}_{\mu \nu} = 0.
\end{equation} 
To find all the solutions, one must pick up a specific form of the background metric.  In the   
TMG case,  all the solutions were constructed in \cite{Strominger} using the $SL(2, R)\times SL(2, R)$ symmetry of $AdS_3$ written as 
\begin{equation}
ds^2 = \ell^2 \Big (-\cosh^2 \rho d\tau^2 + \sinh^2 \rho d\phi^2 + d \rho^2 \Big)
\end{equation}
The solutions will furnish a representation of the algebra and can be built from the primary states given in \cite{Strominger} and remain intact for our case once the viable primary weights (for massive modes) yielding bounded solutions are adjusted as
\begin{eqnarray}
&&(h, \bar  h) = \Big ( \frac{ 3 + p_1 \ell}{2},  \frac{ -1 + p_1 \ell}{2} \Big ) ,\nonumber  \\
&&(h, \bar  h) = \Big ( \frac{ -1 - p_2 \ell}{2},  \frac{ 3 - p_2 \ell}{2} \Big )
\end{eqnarray}
with the conditions that $p_1 \ell \ge 1$ and $p_2 \ell \le -1 $ to ensure  $ h+ \bar h \ge 2$ needed for proper decay of solutions. Of course one also has the usual Einstein modes with the left and right-moving massless gravitons as $(2,0)$ and $(0,2)$, respectively. The primary solutions become
\begin{equation}
h_{\mu \nu} = e^{ -i \tau ( h + \bar h)}  e^{ -i \phi ( h - \bar h)} F_{\mu \nu}(\rho)
\end{equation}
where 
\begin{eqnarray}
F_{\mu\nu}(\rho)=f(\rho)\left(\begin{array}{ccc}
                                         1 & {h-\bar{h}\over2}& {2i\over\sinh(2\rho)} \\
                                          {h-\bar{h}\over2} & 1 & {i(h-\bar{h})\over\sinh(2\rho)} \\
                                         {2i\over\sinh(2 \rho)} &{i(h-\bar{h})\over \sinh(2 \rho)}   & - {4\over\sinh^2(2\rho)} \\
                                       \end{array}\right),
\end{eqnarray}
and 
\begin{equation}
f(\rho)=(\cosh{\rho})^{-(h+\bar{h})}\sinh^2{\rho}.
\end{equation}
All the other solutions of the theory can be constructed from these primary states as descendants using the lowering operators of the algebra.

At the critical point, one has the degeneration ${\mathcal{D}}^{m_1} ={\mathcal{D}}^{L}$ and log-modes appear \cite{Grumiller}, now with the primary weights becoming that of a massless left-moving graviton and a massive graviton with
\begin{equation}
(h, \bar  h) = \Big ( \frac{ -1 +m^2 \ell}{2},  \frac{ 3 + m^2 \ell}{2} \Big )
\end{equation}

Finally, let us compute the energies of the bulk excitations to see that there does not arise negative energies.
Equation for the linearized excitation (\ref{lin_h}) comes from the variation of the action}
\begin{eqnarray}
&I &= -\frac{1}{4} \int \Big (\frac{1}{m^2} (\bar\square h^{\mu \nu})^2  +
 ( 1 -\frac{ 5}{ m^2 \ell^2} )(\bar\nabla_\alpha h_{\mu \nu})^2  \\ \nonumber
& -& \frac{1}{\mu} (\bar\square + \frac{2}{\ell^2})h^{\mu \nu}\eta_{\mu}\,^{\alpha \beta}\bar\nabla_\alpha h_{\beta \nu}  -\frac{2}{\ell^2} (1 - \frac{3}{m^2 \ell^2})h_{\mu \nu}^2 \Big ).
\end{eqnarray}
Ostrogradsky procedure  \cite{tahsin_pope} leads to the following bulk excitation energies (up to positive multiplicative constants) that satisfy the field equations for the massless modes
\begin{equation}
E_{L/R} = - \Big (1 - \frac{1}{m^2 \ell^2} \mp \frac{1}{\mu \ell} \Big )  \int d^3 x \, \sqrt{-\bar{g}} \,\bar{\nabla}^0  h_{L/R}^{\alpha \nu} \partial_t h^{L/R}_{\mu \nu}, 
\end{equation}
and for the massive modes
\begin{equation}
E_{m_i} = - \Big (1 - \frac{1}{m^2 \ell^2}  + \frac{m_i^2}{m^2} \mp \frac{1}{\mu \ell} \Big )  \int d^3 x \,\eta _\alpha\,^{0\mu} h_{m_i}^{\alpha \nu}\partial_t h^{m_i}_{\mu \nu}, \label{energy1}
\end{equation}
The integrals are to be evaluated for all the solutions, which were shown in \cite{Strominger} to yield negative values. Hence all the parentheses should be positive or zero, to have non-ghost excitations and positivity of these terms should not contradict with the positivity of the boundary theory  (\ref{positivity}). This can be achieved if 
\begin{equation}
\mu \ell \ge  \frac{ m^2 \ell^2 }{ (m^2 +m_i^2)  \ell^2  -1},
\end{equation}
which is a weaker condition than (\ref{positivity}) for $m_i^2  \ge 0$. Hence, unlike the case of TMG, one does not have to go to the chiral limit to have a bulk-boundary unitary theory. This is crucial since, not going to the chiral limit, one avoids the problematic log-modes \cite{Grumiller}. 

Finally let us briefly discuss the issue of  "causality" along the lines nicely described in \cite{caus}. Higher derivative terms, such as the ones  introduced here, potentially can bring causality-violating terms. To show that the theory we presented is causal, following \cite{caus}, we here show that the Shapiro time-delay of a test particle traversing  a shock-wave created by a fast moving particle has the correct sign.  As we shall be interested in local causality, flat space considerations are sufficient. Given $T_{uu} = - |P_u| \delta(u) \delta(x)$, to be the energy-momentum  tensor of the fast-moving particle that creates the shock-wave, one has the metric: $ds^2 = - du dv + h(u,x) du^2 + dx^2$ , which solves (\ref{mmg2_denk}) with the above source, if
\begin{equation}
h(u,x)=  \frac{2 m^2 |P_u| \delta(u)}{ p_1^2 - p_2^2}\Bigg( \frac{1}{|p_1|}\theta(x) e^{-|p_1| x}+  \frac{1}{p_2}\theta(-x) e^{p_2 x} \Bigg),
\end{equation}
where for the sake of definiteness we took the $\mu >0, m^2 >0$ case, hence $p_1 = -|p_1|$ , $p_2 >0$ and  $ p_1^2 - p_2^2 >0$ which follow from (\ref{pdenk1}) and (\ref{pdenk2}) and we have set the Newton's constant $\kappa_3=1$
As this is a parity violating theory, the created wave to the left and to the right of the particle differs, but they reproduce  each other once $\mu \rightarrow -\mu$ as expected. So the test particle  with momentum $P_v$  suffers different time delays depending on whether it passes the left or the right of the source particle in the transverse direction. Say,  the  impact 
parameter is $b$, then the time delays are 
\begin{equation}
\Delta v_{x >0}  = \frac{2 m^2 |P_u|  e^{-|p_1| b}}{|p_1|( p_1^2 - p_2^2)}, \,\,\,\Delta v_{x<0}  = \frac{2 m^2 |P_u|  e^{-p_2 b}}{p_2( p_1^2 - p_2^2)},
\end{equation}
both of which are positive, keeping causality intact. This $3D$ result is interesting, since for higher dimensions causality violations in higher derivative theories were shown to be avoided only with infinitely many massive higher spin states \cite{caus}, string theory being the unique example \cite{vene}.
\section{Conclusions}
We have shown that there is a massive spin-2 theory which is free of the bulk and boundary unitarity conflict in 3D.  Moreover, one does not have to go to the chiral gravity limit to satisfy the unitarity and hence no log-modes arise in the generic theory. The theory has two helicity modes and it is quite interestingly related to the single mode theory, that is the Minimal Massive Gravity, which hitherto has  been the only known non-trivial theory free of the bulk-boundary conflict in AdS.
The relation is as follows: New Massive Gravity (NMG), a parity-invariant  massive-spin-2 theory with two massive modes, and which suffers from the bulk-boundary unitarity conflict,  gives birth to two theories, if its field equations are judiciously split, one of which is MMG and the other one is the theory we presented here, which one might perhaps call ``MMG$_2$". 
MMG$_2$   is built on the traceless $H_{\mu \nu}$-tensor that we  defined here and the theory has a unique, an attractive feature since these higher derivative theories typically have more than one vacua which cannot be compared with each other, for example as far as their energy properties are concerned.
Secondly, all the solutions of the theory has constant scalar curvature, which is also a property of TMG.  All the solutions of TMG also solve MMG$_2$ once the topological mass $\mu$ is tuned. Similar solutions inheritance issues from TMG to NMG and other more general theories have been explored in \cite{Aliev,Gurses}. We have not built a Lagrangian for the theory, but following \cite{Townsend1} this can be presumably done with auxiliary fields. 

This work is partially supported by  T\"{U}B\.{I}TAK  grant 113F155. We would like to thank Wout Merbis for a crucial remark regarding the on-shell Bianchi identity of the theory and T.C. Sisman for a critical reading of the paper.


\begin{thebibliography}{10}


\bibitem{DJT} S. Deser, R. Jackiw and S. Templeton, Phys. Rev. Lett.
\textbf{48} 975 (1982); Ann. Phys. (N.Y.) \textbf{140} 372 (1982);
\textbf{185} 406(E) (1988). 

\bibitem{NMG} 
  E.~A.~Bergshoeff, O.~Hohm and P.~K.~Townsend,
  Phys.\ Rev.\ Lett.\  {\bf 102}, 201301 (2009).

\bibitem{Townsend1} E.~Bergshoeff, O.~Hohm, W.~Merbis, A.~J.~Routh
and P.~K.~Townsend, 
 Class.\ Quant.\ Grav.\ \textbf{31},145008 (2014).

\bibitem{Townsend2} 
  A.~S.~Arvanitakis, A.~J.~Routh and P.~K.~Townsend,
  Class.\ Quant.\ Grav.\  {\bf 31}, no. 23, 235012 (2014).

\bibitem{Sinha}
  A.~Sinha,
  JHEP {\bf 1006} 061 (2010).

\bibitem{BINMG}
  I.~Gullu, T.~C.~Sisman and B.~Tekin,
  Class.\ Quant.\ Grav.\  {\bf 27} 162001 (2010).


\bibitem{Strominger} 
  W.~Li, W.~Song and A.~Strominger,
  JHEP {\bf 0804}, 082 (2008).

\bibitem{Tekin} 
  B.~Tekin,
  Phys.\ Rev.\ D {\bf 90}, no. 8, 081701 (2014).

\bibitem{Alishahiha} 
  M.~Alishahiha, M.~M.~Qaemmaqami, A.~Naseh and A.~Shirzad,
  JHEP {\bf 1412}, 033 (2014).

\bibitem{Sun} 
  Y.~Liu and Y.~W.~Sun,
  Phys.\ Rev.\ D {\bf 79}, 126001 (2009).

\bibitem{Setare} 
  M.~R.~Setare,
  arXiv:1412.2151 [hep-th].

\bibitem{pope} 
  D.~D.~K.~Chow, C.~N.~Pope and E.~Sezgin,
  Class.\ Quant.\ Grav.\  {\bf 27}, 105001 (2010).

\bibitem{sezgin} 
  D.~D.~K.~Chow, C.~N.~Pope and E.~Sezgin,
  Class.\ Quant.\ Grav.\  {\bf 27}, 105001 (2010). 

\bibitem{emel} 
  E.~Altas and B.~Tekin, arXiv:1503.04726 [hep-th].


\bibitem{BF} 
  P.~Breitenlohner and D.~Z.~Freedman,
  Phys.\ Lett.\ B {\bf 115}, 197 (1982).

\bibitem{BH} 
  J.~D.~Brown and M.~Henneaux,
  Commun.\ Math.\ Phys.\  {\bf 104}, 207 (1986).

\bibitem{DT1} 
  S.~Deser and B.~Tekin,
  Phys.\ Rev.\ D {\bf 67}, 084009 (2003).

\bibitem{DT2} 
  S.~Deser and B.~Tekin,
  Class.\ Quant.\ Grav.\  {\bf 20}, L259 (2003).

\bibitem{BTZ} 
  M.~Banados, C.~Teitelboim and J.~Zanelli,
  Phys.\ Rev.\ Lett.\  {\bf 69}, 1849 (1992).

\bibitem{Olmez} 
  S.~Olmez, O.~Sarioglu and B.~Tekin,
  Class.\ Quant.\ Grav.\  {\bf 22}, 4355 (2005).

\bibitem{Grumiller} 
  D.~Grumiller and N.~Johansson,
  JHEP {\bf 0807}, 134 (2008).

\bibitem{tahsin_pope} 
  S.~Deser, H.~Liu, H.~Lu, C.~N.~Pope, T.~C.~Sisman and B.~Tekin,
  Phys.\ Rev.\ D {\bf 83}, 061502 (2011).

\bibitem{caus}
  X.~O.~Camanho, J.~D.~Edelstein, J.~Maldacena and A.~Zhiboedov,
  arXiv:1407.5597 [hep-th].

\bibitem{vene} 
  G.~D'Appollonio, P.~Di Vecchia, R.~Russo and G.~Veneziano,
  arXiv:1502.01254 [hep-th].

\bibitem{Aliev} 
  H.~Ahmedov and A.~N.~Aliev,
  Phys.\ Rev.\ Lett.\  {\bf 106}, 021301 (2011).

\bibitem{Gurses} 
  M.~Gurses, T.~C.~Sisman and B.~Tekin,
  Phys.\ Rev.\ D {\bf 86}, 024001 (2012).

\end{thebibliography}
\end{document}